\documentclass[twocolumn,showpacs,aps,pra,superscriptaddress]{revtex4}
\usepackage{graphicx}
\usepackage{bm}
\usepackage{amsmath}
\newcommand{\ignore}[1]{}

\def\beq{\begin{equation}}
\def\eeq{\end{equation}}
\def\la{\langle}
\def\ra{\rangle}

\def\beqa{\begin{eqnarray}}
\def\eeqa{\end{eqnarray}}
\begin{document}
\title{Generalized relation between pulsed and continuous measurements
in the quantum Zeno effect} 
\author{J. G. Muga}
\email{jg.muga@ehu.es}
\affiliation{Departamento de Qu\'\i mica F\'\i sica, UPV-EHU, 
Apdo. 644, 48080 Bilbao, Spain}
\author{J. Echanobe}
\email{javi@we.lc.ehu.es}
\affiliation{Departamento de Electricidad y Electr\'onica, 
UPV-EHU, Apdo. 644, 48080 Bilbao, Spain}
\author{A. del Campo}
\email{adolfo.delcampo@ehu.es}
\affiliation{Departamento de Qu\'\i mica F\'\i sica, UPV-EHU, 
Apdo. 644, 48080 Bilbao, Spain}
\author{I. Lizuain}
\affiliation{Departamento de Qu\'\i mica F\'\i sica, UPV-EHU, 
Apdo. 644, 48080 Bilbao, Spain}
%   
%\author{L. Schulman}
%\email{dwsprung@mcmaster.ca}
%\affiliation{
%New York\\
%  Hamilton, Ontario L8S 4M1 Canada}
%\maketitle
%
\begin{abstract}
A relation is found between pulsed measurements of the
excited state probability of a two-level atom illuminated by a 
driving laser, and a continuous measurement by a second 
laser coupling the excited state to a third state which decays  
rapidly and irreversibly. We find the time between pulses to   
achieve the same average detection time than a given continuous 
measurement in strong, weak, or intermediate coupling regimes,
generalizing the results in 
L. S. Schulman, Phys. Rev. A {\bf 57}, 1509 (1998).    
\end{abstract}
\maketitle

\section{Introduction}
The Quantum Zeno effect has not ceased to attract attention 
since its discovery \cite{Allcock,MS77} due to its importance as a fundamental 
phenomenon and more recently because of different applications, in particular in   
quantum information \cite{Zurek,Hosten,En08} or thermodynamic control \cite{Ku08,Sc08}. 
Basic properties of the effect are being scrutinized both  
theoretically and experimentally, see for recent work \cite{De06,ket06,Liberato,Ec08,Vaidman}
and references therein. One of the relevant and recurrent issues is the
role played by 
the repeated measurements (questioning their necessity \cite{MPS97}) 
and the possibility to 
achieve similar results with continuous measurements.      
The first papers on the quantum Zeno effect 
formulated it as the suppression of transitions between quantum states 
because of continuous observation, {\em mimicked by} 
repeated instantaneous measurements \cite{Allcock,MS77}. This suppression 
was taken as an argument against such a discretized description of the
continuous measurement by Misra and Sudarshan \cite{MS77} but,
in later works, many  authors have emphasized instead the connections 
and similarities between pulsed and continuous models, see \cite{cp1,cp2,Itano,Schulman98,Koshino}
and references therein, in which the transition is 
affected by the measurement in various degrees depending on the 
parameters characterizing the system and the coupling with the measurement
device included in the Hamiltonian.   
Schulman found an elegant relation between discrete and continuous models.
In the simplest case the continuous model represents 
a laser-driven two-level atom with internal states $|1\ra, |2\ra$
subjected to an irreversible measurement of state $|2\ra$ \cite{Schulman98}
with an intense (strong) coupling or quenching laser, see Fig. 1,  or,
equivalently, a fast natural decay of level 2 
which, possibly counterintuitively,  hinder the 1-2 transition in spite of the 1-2
driving laser. (In this work the word ``coupling'' is exclusively associated
with the measurement, 2-3 transition, whereas the ``driving'' refers to the 1-2 transition.)  
In the corresponding pulsed measurement to be compared with the continuous one,  
the 1-2 driving remains but, instead of the continuous measurement via 
effective or natural decay, an instantaneous measurement of level 2 is made every $\delta t$. Schulman noticed that if $\delta t$ is chosen as four times the inverse of the effective decay rate, similar results are obtained in the continuous and pulsed cases, as it has later been observed experimentally \cite{ket06}.     
We shall extend this result by providing a generalized 
connection between the continuous and the pulsed measurements which 
is also valid in the weak coupling regime, in approximate (explicit) or exact
(implicit) forms.

\section{The model}
\subsection{Continuous measurement}

For the continuous model 
we assume a driving laser of frequency $\omega_{L12}$ for the 1-2 transition 
(with transition frequency $\omega_{12}$, and Rabi frequency $\omega$),  and a coupling laser 
of frequency $\omega_{L23}$ for the 2-3 transition (with transition frequency $\omega_{23}$ 
and Rabi frequency $\Omega$). 
The atom in 3 will emit a photon decaying irreversibly outside the original 
1-2 subspace. In a recent experiment this irreversible decay has been realized by the recoil imparted at the photon emission, which gives the atom enough kinetic energy to escape from the confining trap \cite{ket06}. 
We shall adopt a number of standard approximations for the initial Hamiltonian:
semiclassical description for the two lasers, dipole approximation, and rotating-wave approximation, 
\beqa
H&=&\hbar\omega_{12}|2\ra\la 2|+\hbar(\omega_{12}+\omega_{23}-i\Gamma/2)|3\ra\la 3|
\nonumber
\\
&+&
\frac{\hbar}{2}\omega[|2\ra\la 1|e^{-i\omega_{L12} t}+H.c.]
\nonumber
\\
&+&  
\frac{\hbar}{2}\Omega[|3\ra\la 2|e^{-i\omega_{L23} t}+H.c.],
\eeqa
where $H.c.$ means Hermitian conjugate, and $\Gamma$ is the decay rate of level 3.   
In a laser-adapted interaction picture, 
using $H_0=|2\ra\la2| \hbar\omega_{L12}+|3\ra\la3|\hbar(\omega_{L12}+\omega_{L23})$, 
we may eventually write down a time-independent effective Hamiltonian,   
\beq
H_I=\frac{\hbar}{2} \left(
\begin{array}{ccc}
0&  \omega & 0
\\
\omega &-2\delta_0 & \Omega
\\
0& \Omega &-i\Gamma\!-\!2(\Delta+\delta_0)
\end{array}\right).
\label{ham0}
\eeq

%
%%%%%%%%%%%%%%%%%%%%%%%%%%%%%%%%%%%%%%%%%%%%%%%%%%%%%%%%%%%%%%%%%%%%
\begin{figure}[t]
\begin{center}
\includegraphics[width=7.5cm]{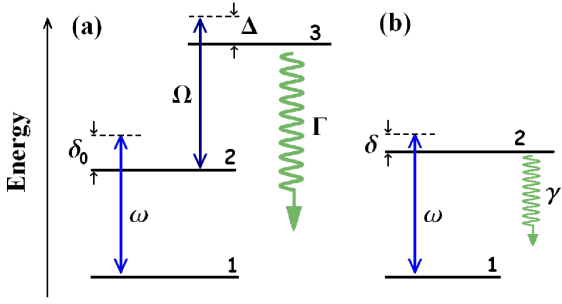}
\end{center}
\caption{\label{ep} (Color online) (a) Energy levels 1, 2, 3; detunings of the two lasers 
with respect to the 1-2 and 2-3 transitions; on-resonance Rabi frequencies, 
and decay rate of level 3; (b) Effective energy scheme 
after the adiabatic elimination of level 3. Note the level shift with respect to 
the scheme in (a).}
\end{figure}
%%%%%%%%%%%%%%%%%%%%%%%%%%%%%%%%%%%%%%%%%%%%%%%%%%%%%%%%%%%%%%%%%%%%%%   
%
% 

%
%where 
%$|1\ra=(100)^T$, $|2\ra=(010)^T$, $|3\ra=(001)^T$, and the subscript
%$T$ means ``transpose''.  
Here  
$\delta_0$ is the detuning (laser frequency  minus transition frequency) 
for the 1-2 transition, and $\Delta$ is the 
detuning for the 2-3 transition.       
For a large enough $\Gamma>>\Omega$,
level 3 is scarcely populated (low saturation)
and it can be adiabatically eliminated to obtain an effective Hamiltonian for the
subspace of levels 1-2,  
\beq
H_I=\frac{\hbar}{2} \left({0\atop \omega}\;\;{\omega
\atop -i\gamma\!-\!2\delta} \right),
\label{ham}
\eeq
where the effective decay rate $\gamma$ 
is related to the measurement coupling parameters \cite{ket06}, 
\beq
\gamma=\frac{\Gamma \Omega^2}{\Gamma^2+4\tilde{\Delta}^2},
\label{ga}
\eeq
and
\beq  
\delta=\delta_0-\tilde{\Delta}\Omega^2/(4\tilde{\Delta}^2+\Gamma^2),
\label{del}
\eeq
with $\tilde{\Delta}=\Delta+\delta_0$. 
To obtain Eqs. (\ref{ga}, \ref{del}) using complex potentials
see \cite{rus04}. 
The general solution of the time-dependent Schr\"odinger equation
corresponding to the 
Hamiltonian (\ref{ham}) is easy to find 
by Laplace transform. 
For the boundary condition $\psi^{(1)}(t=0)=1$, corresponding to the atom being 
initially in the ground state, the ground and excited state 
amplitudes take the form
\beq
\psi^{(1)}=e^{-\gamma_0 t/4}
\left[\cosh(tR/2)+\frac{\gamma_0}{2R}\sinh(tR/2)\right],
\label{psi1}
\eeq
and
\beq
\psi^{(2)}=-i\frac{\omega e^{-\gamma_0 t/4}}{R}\sinh(tR/2),
\label{psi2}
\eeq
where 
\beqa
R&=&(\gamma_0^2-4\omega^2)^{1/2}/2,
\\
\gamma_0&=&-2i\delta+\gamma,  
\eeqa
and the corresponding probabilities $p_{j}$ are given as
$|\psi^{(j)}|^2$, $j=1,2$. 

The detection probability per unit time 
in the continuous measurement 
is $W(t)=\gamma|\psi^{(2)}(t)|^2$, and the  
average lifetime before detection, $\tau_c$,
is given by $\int_0^\infty dt\, t W(t)$. 
Using Eq. (\ref{psi2}),  
\beq
\tau_c=\frac{2}{\gamma}+\frac{\gamma}{\omega^2}
+\frac{4\delta^2}{\gamma\omega^2}.
\label{tauc}
\eeq
%
%{}\vspace*{.2cm}  
%
% 
%%%%%%%%%%%%%%%%%%%%%%%%%%%%%%%%%%%%%%%%%%%%%%%%%%%%%%%%%%%%%%%%%%%%
\begin{figure}[t]
\begin{center}

\vspace{.25cm}
\includegraphics[width=7.cm]{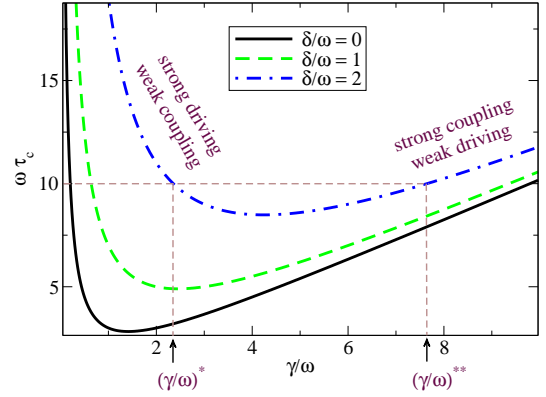}
\end{center}
\caption{\label{czeno} (Color online) Dependence of average lifetime $\tau_c$ of a two-level atom continuously driving by a laser 
(of Rabi frequency $\omega$) 
on the decay rate $\gamma$ for different values of the 
detuning $\delta$. A given value of $\tau_c$ can be generally achieved by
applying weak, $(\gamma/\omega)^*$, or strong coupling,
$(\gamma/\omega)^{**}$.}
\end{figure}
%%%%%%%%%%%%%%%%%%%%%%%%%%%%%%%%%%%%%%%%%%%%%%%%%%%%%%%%%%%%%%%%%%%%%%   
%
%
%%%%%%%%%%%%%%%%%%%%%%%%%%%%%%%%%%%%%%%%%%%%%%%%%%%%%%%%%%%%%%%%%%%%
\begin{figure}[h!]
\begin{center}
\includegraphics[width=7.5cm]{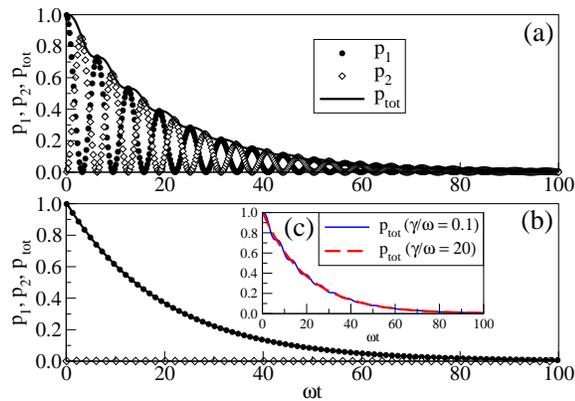}
\end{center}
\caption{\label{czeno2} (Color online) Evolution of the populations in an effective two-level atom
driven by a laser with $\delta=0$, 
in the (a) weak coupling ($\gamma/\omega$=0.1) and (b) strong coupling ($\gamma/\omega=20$) regimes.
Both cases lead to the same lifetime $\tau_c$ and the total populations $p_{tot}=p_1+p_2$ are very similar, 
see (c).}
\end{figure}
%%%%%%%%%%%%%%%%%%%%%%%%%%%%%%%%%%%%%%%%%%%%%%%%%%%%%%%%%%%%%%%%%%%%%%
%
As shown in Fig. \ref{czeno}, the upshot is that the same average lifetime $\tau_c$ can be 
achieved for two different ratios of $\gamma/\omega$, this is, 
both in the strong and weak driving regime. While in the former the system undergoes damped Rabi Oscillations (Fig. \ref{czeno2}a), 
in the later the total probability approximately coincides with that in the ground state $p_{tot}(t)\cong p_{1}$ (Fig. \ref{czeno2}b).
Nonetheless, in both regimes the total probability obeys essentially the same decay law (Fig. \ref{czeno2}c).

\subsection{Pulsed measurement}
Alternatively, a pulsed measurement of the population at 2  may be 
performed on the 
same two-level atom described before,
which evolves between measurements with Hamiltonian   
\beq
H_I(\Omega=0)=\frac{\hbar}{2} \left({0\atop \omega}{\omega
\atop -2\delta_0} \right),  
\label{ham2}
\eeq
equal to (\ref{ham}) when $\Omega=0$. 
Note the level shift with respect to the Hamiltonian (\ref{ham})
because of the absence of the
2-3 coupling.   

At intervals $\delta t$, the population in 2 is measured ``instantly''
(in practice, in a time scale $\tau_m$ small compared to $\delta t$, but we shall ignore 
here the corrections of order $\tau_m/\delta_t$, see e.g. \cite{ket06}.)
and removed from the atomic ensemble.  
This process is represented by successive projections onto the 
state $|1\ra$ at times $\delta t, 2\delta t,...,k\delta t,..$.   
%
%%%%%%%%%%%%%%%%%%%%%%%%%%%%%%%%%%%%%%%%%%%%%%%%%%%%%%%%%%%%%%%%%%%%
\begin{figure}[h!]
\begin{center}
\includegraphics[width=7.5cm]{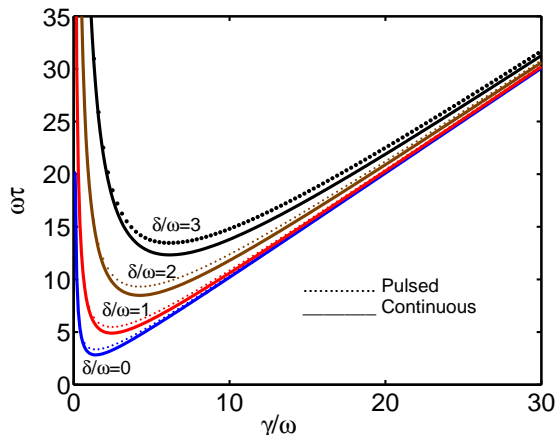}
\end{center}
\caption{\label{ep} (Color online) (a) Lifetimes before detection for continuous ($\tau=\tau_c$, solid lines) 
and pulsed ($\tau=\la t\ra$ using Eqs. (\ref{exactp}) and (\ref{dt}), dashed lines) measurements for different values of detuning.
For the pulsed calculations we have taken $\delta_0=\delta$ corresponding to $\tilde{\Delta}=0$.}
\end{figure}
%%%%%%%%%%%%%%%%%%%%%%%%%%%%%%%%%%%%%%%%%%%%%%%%%%%%%%%%%%%%%%%%%%%%
\begin{figure}[h!]
\begin{center}
\includegraphics[width=7.5cm]{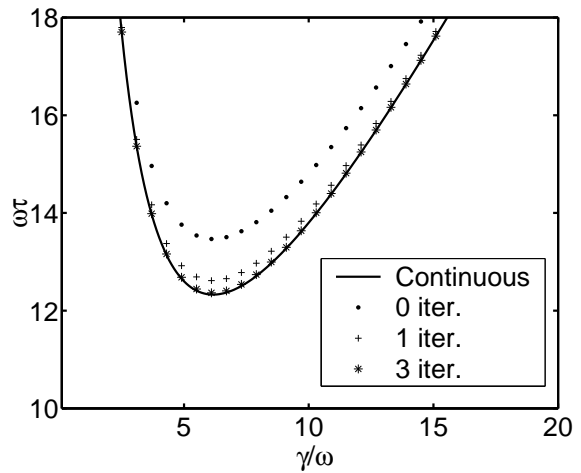}
\end{center}
\caption{\label{epb} Lifetimes before detection for $\delta/\omega=3$.
Solid and dotted lines
are, as in Fig. 4, for the continuous measurement 
and for pulsed measurement, using (\ref{dt}), respectively.
The symbols represent the first and third iteration of Eq. (\ref{iter}), 
starting with the seed of Eq. (\ref{dt}).}
\end{figure}
%%%%%%%%%%%%%%%%%%%%%%%%%%%%%%%%%%%%%%%%%%%%%%%%%%%%%%%%%%%%%%%%%%%%%%
%The solution for the Hamiltonian (\ref{ham2}) is obtained by 
%making $\Omega=0$. 
%
The probability to detect the atom in the $k$-th measurement instant is the probability to 
find the atom in level 2 at that instant, 
$P_2 (1-P_2)^{k-1}$, 
where we have introduced $P_2$, to be distinguished from the more generic $p_2$,  as the probability to find the excited state 
at $\delta t$ without coupling,   
\beq
P_2\equiv|\psi^{(2)}(\Omega=0,t=\delta t)|^2,
\eeq
when the (normalized) state is in the ground state at $t=0$. 
The average detection time is thus calculated, exactly, as
\beqa
\la t\ra&=&\sum_{k=0}^\infty k \delta t P_2(1-P_2)^{k-1}=
\frac{\delta t P_2}{1-P_2}\sum_{k=0}^\infty k(1-P_2)^k
\nonumber\\
&=&
\delta t/P_2,
\label{exactp}
\eeqa
where the last sum can be performed by taking the derivative of the
geometric series. 
\subsection{Connection between continuous and pulsed measurements}
The objective here is to find the time between pulses, $\delta t$, so that 
the average detection time in the pulsed measurement, $\la t\ra$,  equals 
the lifetime $\tau_c$ of a given continuous measurement.   
Let us first pay attention to the short-time 
asymptotics for $H_I(\Omega=0)$. From Eq. (\ref{psi2}), with $\Omega=0$, 
the dominant term for the probability of level 2 at $\delta t$, $P_2$,
is independent of $\delta_0$ (a remarkable
result that only holds at short times), and takes the 
simple form
\beq
P_2\sim \delta t^2\omega^2/4=(\delta t/\tau_Z)^2.
\label{p2a}
\eeq
The short-time dependence is characterized by a ``Zeno time''
$\tau_Z\equiv 2/\omega$ which defines the time-scale for the quadratic time 
dependence \cite{Schulman98}.
Equating $\tau_c$ and  
$\la t\ra$, 
see Eqs. (\ref{tauc}) and (\ref{exactp}),
with the quadratic approximation (\ref{p2a}) for $P_2$ 
we get   
\beq
\delta t=\frac{4\gamma}{2\omega^2+\gamma^2+4\delta^2}. 
\label{dt}
\eeq
The same result may be obtained ``\`a la Schulman'' \cite{Schulman98} as follows: 
after $k$ successive pulses separated by $\delta t<<\tau_Z$, at $t=k\delta t$,
the probability of the ground state (which is the total remaining probability 
after removing $p_2$ in each pulse) will assume 
the form 
\beq
p_1(t)\approx\left(1-\frac{t}{k}\frac{\delta t}{\tau_Z^2}\right)^k\approx\exp(-t/\tau_{EP}),
\label{2aprox}
\eeq
i.e., an effective exponential decay law $\exp(-t/\tau_{EP})$ where  
$\tau_{EP}=\tau_Z^2/\delta t$ is the effective lifetime.  
By equating $\tau_{EP}$ and $\tau_c$ we get again Eq. (\ref{dt}). 
This requires $\delta t<<2/\omega$ or $\tau_Z<<\tau_c$.
%This is one of the main results of this work.      
These inequalities and therefore the exponential approximation
are well fulfilled for weak 1-2 driving, $\omega/\gamma<<1$ (corresponding to strong 2-3 coupling), 
or strong 1-2 driving conditions $\omega/\gamma>>1$ (corresponding to weak 
2-3 coupling), but not so well for intermediate driving conditions, see the region near the 
minimum at $\gamma/\omega=[2+4(\delta/\omega)^2]^{1/2}$ in Fig. \ref{ep}. 
For a more accurate treatment we may find the really optimal 
$\delta t$ by imposing the equality between the exact expressions for the 
lifetimes, $\tau_c=\la t\ra$, and solving the implicit 
equation (\ref{exactp}), 
e.g. by iteration, 
\beq
\delta t=\tau_c P_2(\delta t). 
\label{exact}
\eeq
This may be carried out by a Newton-Raphson method,
%  e solution of Eq. (\ref{exact}) represents the crossing of the straight line %$\delta t$ and the curve $\tau_c P_2(\delta_t)$. Linearizing $P_2$ at the crossing 
%region as $P_2(\delta t)=P_2(\delta t_0)+P'_2(\delta t_0)(\delta t-\delta t_0)$,
%where $\delta t_0$ is some approximate value for the crossing point, and the prime %means derivative with respect to $\delta t$, 
%we obtain the iterative recipe    
%
\beq
\delta t=\frac{P_2(\delta t_0)-\delta t_0 P_2'(\delta t_0)}{\tau_c^{-1}- P_2' (\delta t_0)}, 
\label{iter}
\eeq
where the $\delta t$ calculated on the left hand side 
becomes the new $\delta t_0$ for the following iteration
and the prime means derivative with respect to $\delta t$. 
A good seed for the first $\delta t_0$ may be Eq. (\ref{dt}), 
and $P_2'$ is very well approximated by $\delta t_0 \omega^2/2$.  
Applying Eq. (\ref{iter}) just once in this way 
provides a very good, approximate, but explicit expression of $\delta t$, 
and the corresponding $\la t\ra$, Eq. (\ref{exactp}), 
is quite close to the continuous measurement result $\tau_c$. Fig. \ref{epb}
is a closeup of one of the cases in Fig. \ref{ep} and depicts a 
few iterations. Three of them are enough to produce a 
$\la t\ra$  indistinguishable from $\tau_c$ in the scale of the figure.        
\section{Discussion}
From Eq. (\ref{tauc}) we see that the average detection time 
in the continuous measurement may be large both for  
strong measurement coupling and in the opposite weak-coupling case.  
A given lifetime in a Zeno-type pulsed measurement may thus be 
reproduced with continuous measurement by means of different conditions. For a fixed detuning $\delta$ and Rabi frequency, these are generally realized with two different values of $\gamma$ 
%While only the strong coupling case (for the larger $\gamma$ value) shares with the
%pulsed Zeno effect the hindering effect of a strong action on the system, 
%it is remarkable that the same effect, at least regarding the average detection time, 
%can be achieved and predicted for a particular weak coupling. Note that the %time-dependent evolutions are not the same: the pulsed evolution is closer to the %continuous evolution with strong coupling, which, unlike the weak-coupling 
%case, does not present Rabi oscillations in the individual probabilities $p_{1,2}$. 
for which the sum $p_1+p_2$ (probability for the atom to remain undetected) is quite similar. This may have practical consequences as a control mechanism,  since a strong continuous coupling could be difficult to produce or cause some undesirable effects. For example, if we want
to maintain the effective two-level scenario, $\Omega$ cannot
be made arbitrarily large in the scale of $\Gamma$ to avoid a
significant population of level 3 which would make the adiabatic
elimination invalid. One more problem could be the excitation of 
other levels, breaking down, again, the simple two (or even three)
level picture.   
    
%%%%%%%%%%%%%%%%%%%%%%%%%%%%%%%%%%%%%%%%%%%%%%%%%%%%%%%%%%%%%%%%%%%%
\begin{figure}[t]
\begin{center}
\includegraphics[width=7.5cm]{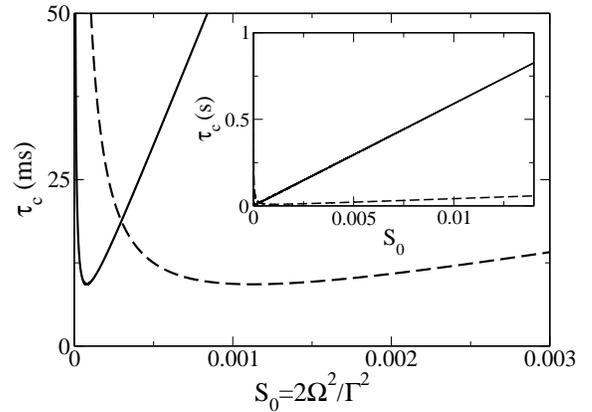}
\end{center}
\caption{\label{ex} Lifetimes before detection versus the 
saturation parameter $s_0=2\Omega^2/\Gamma^2$. 
$\omega=2 \pi\times 48.5$ Hz, $\Gamma =2\pi\times 1.74$ MHz. 
Solid line: $\Delta=\delta_0=\delta=0$; Dashed line: $\Delta=
2\pi\times 3.18$ MHz, and $\delta_0$ is adjusted for each $s_0$ 
so that $\delta=0$.}
\end{figure}
%%%%%%%%%%%%%%%%%%%%%%%%%%%%%%%%%%%%%%%%%%%%%%%%%%%%%%%%%%%%%%%%%%%%%% 
We have in summary generalized Schulman's relation between pulsed and continuous measurement in a two-level system in different ways: Schulman's result corresponds to 
the case $\delta=0$, with $\gamma/\omega>>1$, for which $\delta t=4/\gamma$. 
Our relation (\ref{dt}) is not restricted to these conditions and is generally applicable. It is, however, an approximate one, since it relies on an  
approximate expression for the average measurement time in the pulsed case. 
We have also 
provided the exact result, by solving the implicit equation (\ref{exact}). 
The present results should not be difficult to verify experimentally  
with very similar settings as the ones already used in  
\cite{ket06}. Fig. \ref{ex} shows the average 
continuous time $\tau_c$ versus the
saturation $s_0=2\Omega^2/\Gamma^2$ using parameters from
\cite{ket06}, see the caption for details.
The solid line is for the case in which the detunings of the two
transitions are set to zero, namely
$\delta=\Delta=\delta_0=0$. 
%Also shown are symbols in which the integration time 
%$t_m$ in $ \int_0^{t_m} t W(t)dt/\int_0^{t_m}$ has been limited to 2 seconds.  
%The deviation from the straight 
%line at the higher $s0$ values is similar to the experimental one and 
%emphasizes the increasing importance of the delicate long-time tails
%of the distribution $W(t)$ with $s_0$.      
In the dashed line the measuring laser
detuning is fixed to 
$\Delta=-20\times10^6$ s$^{-1}$, and $\delta_0$ is
adjusted for each measuring laser 
intensity (i.e., for each value of $s_0$) so that $\delta=0$. In principle this requires to solve a qubic equation for $\delta_0$,
see Eq. (\ref{del}), but an excellent 
approximation for the parameter values in \cite{ket06} 
(low saturation) is obtained by taking $\tilde{\Delta}\approx \Delta$, 
\beq
\delta_0(\delta=0)\approx\frac{\Delta \Omega^2}{4 \Delta^2+\Gamma^2}.
\eeq
In practice $\delta_0(\delta=0)$ may be found  by minimizing $\tau_c$
(or maximizing the reduction of the population of $|1\ra$ \cite{ket06})
with respect to the rf frequency $\omega_{12}$. We plot this case to show
that the weak driving region is accessible with parameters 
very near the ones used in \cite{ket06}. 
The deviation from the straight line is clearly 
noticeable around and below $s_0=.001$ which corresponds 
in \cite{ket06} to an optical power of 
$0.83$ $\mu$W.   
In all cases we have solved the dynamics of the three-level system 
numerically and of the effective two-level system analytically 
without finding any disagreement between the two models 
in the scale of the figure.

%or previous experiments \cite{...}.         
%
%
\begin{acknowledgments}
The authors acknowledge discussions with L. Schulman.  
This work has been supported by Ministerio de Educaci\'on y Ciencia (BFM2003-01003), and UPV-EHU (00039.310-15968/2004). A. C. acknowledges financial support by the Basque Government (BFI04.479).
\end{acknowledgments}
%
%
%\begin{references}

\end{document}